\documentclass{article}
\usepackage{spconf,amsmath,graphicx}
\usepackage{amssymb, mathtools}
\usepackage{siunitx}
\usepackage{enumitem}
\usepackage{placeins}
\usepackage{cleveref}
\usepackage{booktabs}
\usepackage{tabularx}
\usepackage{bm}
\usepackage{jabbrv}
\usepackage[numbers, sort&compress, square]{natbib}
\usepackage{tabu}
\usepackage{lipsum}
\usepackage{color}
\usepackage{url}
\usepackage{fancyhdr}

\renewcommand*{\b}{\bfseries}

\usepackage{xcolor}
\iftrue 
    \newcommand{\tho}[1]{\noindent}
\else
    \newcommand{\tho}[1]{\textcolor{blue}{#1}}
\fi

\newcommand{\xtho}[1]{\noindent}

\usepackage{comment}

\DefineJournalException{Proceedings of the IEEE International Conference on Acoustics, Speech and Signal Processing}{Proc. ICASSP}
\DefineJournalException{Proceedings of the Annual Conference of the International Speech Communication Association}{Proc. Interspeech}
\DefineJournalException{IEEE/ACM Transactions on Audio Speech and Language Processing}{IEEE/ACM Trans. ASLP}
\DefineJournalException{IEEE Transactions on Audio Speech and Language Processing}{IEEE Trans. ASLP}
\DefineJournalException{Proceedings of the IEEE/CVF Conference on Computer Vision and Pattern Recognition}{Proc. CVPR}
\DefineJournalException{Proceedings of the 3rd International Conference on Learning Representations}{Proc. ICLR}

\title{End-to-End Complex-Valued Multidilated Convolutional Neural Network for Joint Acoustic Echo Cancellation and Noise Suppression}
\name{
    Karn N. Watcharasupat$^{\star}$,
    Thi Ngoc Tho Nguyen$^{\star}$, 
    Woon-Seng Gan$^{\star}$,
    Shengkui Zhao$^{\dagger}$,
    and Bin Ma$^{\dagger}$
\thanks{%
    This work was supported by the Alibaba Group and the Alibaba-NTU Singapore Joint Research Institute via the Alibaba Innovative Research (AIR) Program (Ref. AN-GC-2020-015).
    K. N. Watcharasupat acknowledges the support from the CN Yang Scholars Programme, NTU.
}
}
\address{%
    $^{\star}$%
    School of Electrical and Electronic Engineering, 
    Nanyang Technological University (NTU), Singapore\\
    $^{\dagger}$%
    Alibaba Group\\
    \{karn001, nguyenth003, ewsgan\}@ntu.edu.sg,
    \{shengkui.zhao, b.ma\}@alibaba-inc.com.
}

\begin{document}
\ninept
\maketitle
\begin{abstract}

Echo and noise suppression is an integral part of a full-duplex communication system. Many recent acoustic echo cancellation (AEC) systems rely on a separate adaptive filtering module for linear echo suppression and a neural module for residual echo suppression. However, in practice, adaptive filtering modules require time to converge and remain susceptible to changes in the acoustic environment. This introduces unnecessary delays to AEC systems using this two-stage framework, despite neural modules already having the capability to suppress both linear and nonlinear echo components. In this paper, we exploit the offset-compensating property of complex time-frequency masks and propose an end-to-end complex-valued neural network architecture. The building block of the proposed model is a pseudocomplex extension of the densely-connected multidilated DenseNet (D3Net), resulting in a very small network of only 354K parameters. The architecture utilized the multi-resolution nature of the D3Net to eliminate the need for pooling, allowing feature extraction using large receptive fields without any loss of output resolution. We also propose a dual-mask technique for joint echo and noise suppression with simultaneous speech enhancement. Evaluation on both synthetic and real test sets demonstrated promising results across multiple energy-based metrics and perceptual proxies.

\end{abstract}
\section{Introduction}
\label{sec:intro}

Noises and echoes are well-known sources of degradation for speech quality in a full-duplex communication (FDC) system. In a FDC system, several speech and noise sources can be at play at each end of the communication line, in addition to the echo signal incident on the nearend microphone due to the playback of the farend loopback signal. Consider a realistic scenario with two desired speech sources, one nearend and one farend, with noise sources on each end. The input signal to the nearend microphone is given by
\begin{equation}
    p[t] = s[t] + n[t] + f(q)[t], 
\end{equation}
where $t$ is the time index, 
$q[t] = r[t] + m[t]$ is the farend microphone input; 
$f(\cdot)$ is a distortion function responsible for both linear and nonlinear transformations in the communication system; $s[t]$ and $r[t]$ are speech sources from the nearend and farend speakers, respectively; 
and, $n[t]$ and $m[t]$ are noise sources from the nearend and farend environments, respectively. For simplicity, we absorb any linear time-invariant transformation of the sound sources due to the environment and the sensors into the signal terms. We denote the complex-valued time-frequency (TF) domain counterpart for each signal by its corresponding boldface uppercase, e.g., $\mathbf{P}[\omega,k]$ for $p[t]$, with $\omega$ and $k$ being the frequency and frame indices, respectively.

In traditional acoustic echo cancellation (AEC) that only utilized linear filtering techniques, noise terms are often ignored or absorbed into the speech terms. Hence, the signal models are often viewed as some variants of
\begin{equation}
    \mathbf{P}[\omega,k] = \mathbf{S}[\omega,k] + \mathbf{W}[\omega,k] \mathbf{Q}[\omega,k] + g(\mathbf{Q}[\omega,k]),
\end{equation}
with the goal of estimating $\mathbf{W}^{{-1}}[\omega,k]$ in either time or TF domain to remove the linearly transformed component of the farend signal. The term $g(\mathbf{Q}[\omega,k])$ is often called residual echo term, which captures any nonlinear components that are not removable by the linear AEC system. Many recent AEC systems have adopted a two-stage approach with a traditional digital signal processing (DSP) module for linear AEC followed by a neural network module for residual echo suppression (RES) \cite{Valin2021Low-ComplexityPercepnet, Peng2021ICASSPSuppression, Halimeh2021CombiningCancellation, Ivry2021DeepSuppression, Wang2021WeightedAEC-Challenge, Pfeifenberger2021AcousticLearning, Seidel2021Y-NetSuppression, Peng2021AcousticInformation}.

Neural RES is highly related to the task of deep noise suppression (DNS), the latter focusing on noise removal rather than echo cancellation. In a broad sense, the residual echoes can be seen as a noise source, thus a few works have adapted DNS models as the RES module for AEC systems \cite{Sridhar2021ICASSPResults, Braun2021InterspeechChallenge}. Given that AEC, RES, and DNS share more similarities than differences, a more unifying formulation naturally follows. Both speech enhancement tasks share a common goal of retrieving the clean speech term $s$, with the differences mainly lying in the sources of degradation. Moreover, neural networks are often capable of performing both linear and nonlinear filtering, thus diminishing the overall benefits of having a linear DSP preprocessor. In particular, modules such as adaptive filters and delay compensators require a `warm-up' period for convergence and can be susceptible to changes in the acoustic environments \cite{Braun2021InterspeechChallenge}, especially in a doubletalk scenario. On the other hand, neural modules, once trained, operate on a fixed set of weights and provide consistent enhancement without the need for an initial convergence period. 

With these considerations in mind, we propose an end-to-end joint noise and echo suppression model based on the D3Net building blocks \cite{Takahashi2021DenselyTasks}. Inspired by \cite{Hu2020DCCRN:Enhancement}, we introduce a pseudocomplex extension of the D3Net, allowing fast computation even on hardware without native complex number supports. Unlike most systems based on convolutional neural networks (CNNs), the proposed architecture exploits the multi-resolution nature of the D3Net building blocks, and operates without any pooling, preserving full feature resolutions throughout the network.

We further proposed a dual-masking technique using two complex-valued output masks to allow the model to simultaneously perform echo/noise suppression and speech enhancement. The complex-valued nature of the masks allows the model to natively compensate delays between the loopback and the microphone signals, eliminating the need for an additional delay compensation module. Due to the purely convolutional nature, the model, with only about 354K parameters, can be easily adapted for causal real-time implementation in future work.

\section{Proposed System}

\subsection{Complex Densely-Connected Multidilated DenseNet}

The densely-connected multidilated DenseNet (D3Net) was formally introduced in \cite{Takahashi2021DenselyTasks} as a building block of a U-Net architecture for semantic segmentation and music source separation, both achieving state-of-the-art performances. Variations of the D3Net have also been used successfully with near state-of-the-art performances on sound event localization and detection \cite{shimada2021accdoa, Shimada2021EnsembleDetection}. 

Although the U-Net structure has been shown to be very effective in source separation, noise suppression, and speech enhancement, it is notorious for being extremely computationally expensive and can rarely be adapted for real-time implementation. Moreover, while the pooling operations have been useful in non-dilated CNNs to capture features at different scales, the multidilated nature of the D3Net is already capable of capturing these scale-dependent features without the need for pooling, which is prone to poor output resolutions. As a result, the proposed model was designed in a sequential manner to ensure adaptability for real-time implementation, and without pooling to preserve feature resolutions. \xtho{this sentence is a bit dangerous, pooling is not the main problem, we replace pooling with dilation, which still have problem of look ahead}

In this work, we adapted the D3Net building blocks described in \cite{Takahashi2021DenselyTasks} by extending them to the complex-valued domain using the pseudocomplex technique in \cite{Hu2020DCCRN:Enhancement}. 
For a layer $\mathcal{H}$ with trainable parameters, such as the convolutional layer, a pseudocomplex extension is given by
\begin{equation}
    \tilde{\mathcal{H}}(\mathbf{Z}) = \left[\mathcal{H}_\text{R}\left(\Re \mathbf{Z}\right) - \mathcal{H}_\text{R}\left(\Im \mathbf{Z}\right)\right] + \jmath\left[\mathcal{H}_\text{I}\left(\Im \mathbf{Z}\right) + \mathcal{H}_\text{I}\left(\Im \mathbf{Z}\right)\right],
\end{equation}
where $\mathcal{H}_\text{R}, \mathcal{H}_\text{I}$ are independent layers of identical functions. 
For a parameterless operation $h$, such as activations, the pseudocomplex extension $\tilde{h}$ is simply $h$ applied on the real and imaginary parts separately, that is,
\begin{equation}
    \tilde{h}(z) = h\left(\Re z\right) + \jmath h\left(\Im z\right).
\end{equation}

\begin{figure}[th]
    \centering
    \includegraphics[width=0.9\columnwidth]{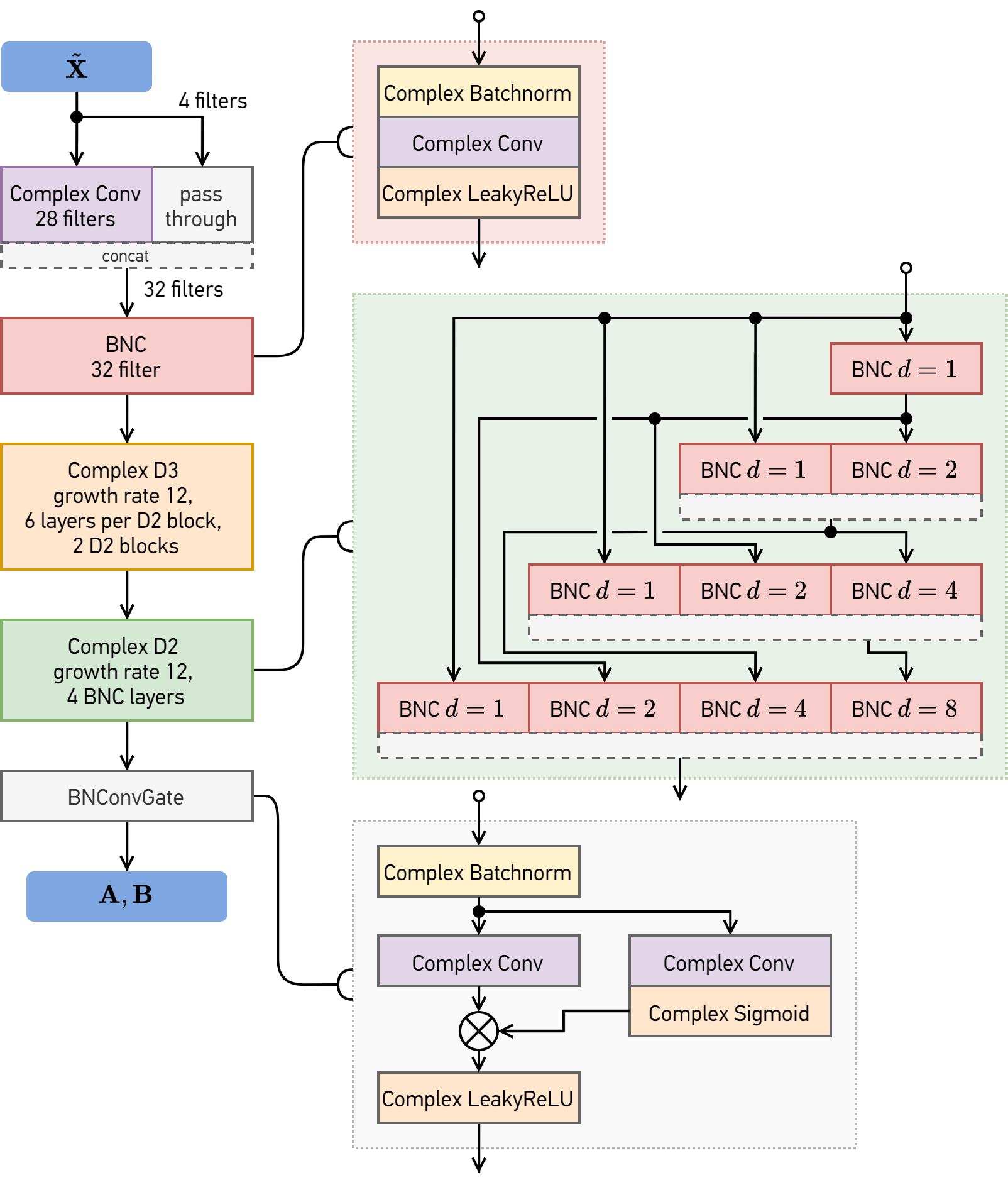}
    \vspace{-10pt}
    \caption{cD3Net Architecture. BNC: Batch normalization-convolutional block. D2: multidilated DenseNet block. D3: densely-connected D2 block.}
    \label{fig:network}
\end{figure}

The overall architecture, the complex densely-connected multidilated DenseNet (cD3Net), is shown in \Cref{fig:network}. We replaced the ReLU activations in the original model with leaky ReLU to ensure the output masks can take any value from the entire complex space. The pseudocomplex extension was applied at the layer level, i.e., all convolutional layers are individually pseudocomplexified. Since the pseudocomplex extensions are real-valued modules under the hood, training and inference of the model can be done efficiently even on hardware without native complex number support.

\subsection{Input Representation}

The complex-valued TF-domain nearend microphone signal $\mathbf{P}$ and the farend loopback signal $\mathbf{Q}$ are used as the input signals. 
%
%
The input tensor was formed via channel-wise stacking of the TF-domain input signals such that
\begin{equation}
    \tilde{\mathbf{X}} = \text{Concat}\left[
        \mathbf{P}, \mathbf{Q}, \mathbf{P}+\mathbf{Q}, \mathbf{P}-\mathbf{Q}
    \right] \in\mathbb{C}^{F\times K\times 4}.
\end{equation}
The latter two channels were explicitly provided to the model to facilitate the implicit learning of the distortion function $f$ required to generate the echo cancellation masks. The latter two channels also provide temporal cues to the model, especially when there is a non-negligible time offset between the echo and the loopback signals.

\subsection{Time-Frequency Masking}

Conventionally, the most popular TF masking method used in AEC, DNS, and several other speech enhancement tasks, has been to apply a mask $\mathbf{A}\in\mathbb{C}^{T\times F}$ (or $\mathbb{R}^{T\times F})$ on the nearend microphone signal ${\mathbf{P}\in\mathbb{C}^{T\times F}}$ to obtain an enhanced speech signal $\widehat{\mathbf{S}} = \mathbf{A} \circ \mathbf{P}$,
where $\circ$ is the elementwise multiplication operator.
\xtho{explain the notation $\circ$}Although this method works very well in a single-track enhancement, this masking technique would discard the rich acoustic information already available in the farend loopback signal for echo suppression.

Since the farend signal is available, we consider the use of two separate masks, $\mathbf{A}, \mathbf{B}\in\mathbb{C}^{T\times F}$, where $\mathbf{B}$ is mainly responsible for echo suppression, and $\mathbf{A}$ for noise suppression and speech enhancement. The dual masking operation is given by 
\begin{equation}
    \widehat{\mathbf{S}} = \mathbf{A}\circ\left(\mathbf{P}  - \mathbf{B}\circ\mathbf{Q}\right).
\end{equation}
In particular, this dual masking scheme is inspired by the traditional cascade of echo and noise suppressors, but adapted for a unified joint enhancement network. Since $\mathbf{B}$ is complex-valued, when the misalignment between the farend loopback and the nearend microphone signal is not significantly greater than the short-time Fourier transform (STFT) window size, there is a reduced need for time-domain compensation as $\mathbf{B}$ itself can partially account for the delay operation.

\section{Experiments}
\subsection{Training}

All models in this paper were trained using only the synthetic dataset of the Microsoft AEC Challenges \cite{Sridhar2021ICASSPResults, Braun2021InterspeechChallenge}. The synthetic dataset provides four signals for each acoustic scene: clean nearend speech signal, potentially noisy farend loopback signal, potentially noisy echo signal, and potentially noisy nearend microphone signal (a mixture of nearend speech, nearend noise, and echo signal). We used the nearend microphone and the farend loopback signals as inputs. The clean nearend speech signals used as the training targets were first scaled to the same amplitude as the speech content in the nearend microphone signal, using the scaling factors provided in the ground truths. The provided echo signals were not used in this paper. The official train and test set consists of 9.5K and 500 acoustic scenes, respectively. We take another 500 samples out of the train set to form a validation set, leaving 9K acoustic scenes for training.

All models in this paper were trained for \num{75} epochs each. Each epoch drew a random subset of \num{2048} scenes from the train set. All models were trained with an effective batch size of \num{32} using an Adam optimizer \cite{Kingma2014Adam:Optimization} with a learning rate of \num{e-3}, weight decay of \num{e-6}, and learning rate reduction by a factor of \num{0.9} for every three epochs without validation loss improvement. Since the audio signals were provided at \SI{16}{\kilo\hertz} sampling rate, we used a 512-sample (\SI{32}{\milli\second}) square-root Hann window with \SI{50}{\percent} overlaps for STFT.

\subsection{Loss Function}

In order to balance between echo and noise cancellation in an energy-based sense and speech enhancement in a perceptual sense, we consider the use of a weighted loss function between the negative signal-to-distortion ratio loss and the perceptual PMSQE loss \cite{Martin-Donas2018AQuality}, given by 
\begin{equation}
    \mathcal{L}(\hat{s}; s) = \alpha\cdot\text{NegSDR}(\hat{s}; s) + \beta\cdot\text{PMSQE}(\hat{s}; s),
\end{equation}
where $\alpha, \beta$ are weight terms which sum to one.

Unlike most systems utilizing SDR-based losses \cite{Kolbaek2020OnEnhancement}, the negative SDR loss is implemented using the scale-dependent SDR (SD-SDR), instead of the scale-invariant counterpart (SI-SDR)\footnote{We have repeated the reported experiments with the SDR and SI-SDR losses replacing the SD-SDR loss; SD-SDR loss performed best across all metrics. Full experimental details were omitted due to space constraints.}, in order to encourage the model to preserve amplitude scaling of the clean speech estimates \cite{LeRoux2019SDRDone}, especially since the amplitude scaling information between the clean speech and the nearend microphone signals is available. For zero-mean estimates and targets, the negative SD-SDR loss is given by
\begin{equation}
    \text{NegSDR}(\hat{s}; s) 
    = -10\log_{10}\left[
        \textstyle\dfrac{\sum_t \left(\gamma s[t]\right)^2}{\sum_t \left(\hat{s}[t]-s[t]\right)^2}
    \right],
\end{equation}
where $\gamma = {\sum_{\tau} \left(\hat{s}[\tau]s[\tau]\right)}/{\sum_{\tau} (s^2[\tau])}$.


\subsection{Data Augmentation}
\label{ssec:daug}

We experimented with a few data augmentation techniques\footnote{We also experimented with additional noise and codec companding, but these techniques did not result in appreciable performance gains.} in the raw signal domain to allow the model to be more robust to realistic signal degradation in a FDC system. For this work, the augmentation was always applied to the farend loopback signal since not all components of the signals that constitute the nearend microphone signal were available in the dataset. Whenever activated, each augmentation technique was applied with a \SI{50}{\percent} probability independently of one another. 

Specifically, we experimented with time shifting between the farend loopback signal and the nearend microphone signal to simulate asynchronous signal arrivals, and amplitude scaling to allow the model to be more robust to discrepancy between the amplitudes of the audio content in the loopback signal and the echo signal. A time shift of up to $\pm\num{512}$ samples, which is the STFT block size used in this paper, was applied on the farend loopback signal. Amplitude scaling of a factor between \num{0.5} to \num{1.5} was applied to the farend loopback signal.






\section{Results and Discussion}

\subsection{Baseline Models}

For comparison, we adapted NSNet \cite{Xia2020WeightedEnhancement}, Conv-TasNet \cite{Luo2019Conv-TasNet:Separation}, and DCCRN \cite{Hu2020DCCRN:Enhancement} for AEC. All three models have their first layers modified to support a multichannel input consisting of the nearend microphone and farend loopback signals. NSNet was additionally modified to use the same STFT parameters as the cD3Net. All other settings of DCCRN and Conv-Tasnet followed the defaults provided by Asteroid \cite{Pariente2020Asteroid:Researchers}. All baseline models used the traditional single-mask method and were trained using the same amount of data and epochs as the proposed models. 

\subsection{Evaluation}

We report two sets of results for each model: one on the test split of the synthetic dataset \cite{Sridhar2021ICASSPResults} used for the development, and one on the Interspeech 2021 AEC Challenge blind test set \cite{Braun2021InterspeechChallenge}. The blind test set consists of 800 real-world recordings with challenging practical conditions, e.g., echo path changes, clock drifts, glitches, noises, gain variations, and onboard DSP.

For the synthetic test set, we evaluate the performance of the models based on four objective metrics commonly used in AEC: scale-invariant source-to-distortion ratio (SI-SDR) \cite{LeRoux2019SDRDone}, standard source-to-distortion ratio (SDR) \cite{Vincent2006PerformanceSeparation}, short-time objective intelligibility (STOI) \cite{Taal2010ASpeech}, and perceptual evaluation of speech quality (PESQ) \cite{Rix2001PerceptualCodecs}. 

For the blind real test set, we evaluate the model using objective proxies of mean opinion score (MOS) via the DNSMOS system \cite{Reddy2020TheFramework} and degradation MOS via the AECMOS system \cite{Cutler2021CrowdsourcingImpairment,  Braun2021InterspeechChallenge}. The AECMOS provides two degradation MOS (DMOS) scores. The Echo DMOS score captures degradation due to farend echo while the Other DMOS captures degradation due to any other sources. Both DMOS are rated on a \num{1}-to-\num{5} scale, with a higher score reflecting a better audio quality. As per the convention in the ICASSP and Interspeech AEC Challenges \cite{Sridhar2021ICASSPResults, Braun2021InterspeechChallenge}, we report MOS for the nearend singletalk scenario, Echo DMOS for the farend singletalk scenario, and both Echo and Other DMOS for the doubletalk scenario.

\renewcommand{\arraystretch}{0.8}
\begin{table*}[t]
    \centering
    \footnotesize
    \begin{tabularx}{\textwidth}{
        Xr
        *{2}l
        l
        *{8}{r}
    }
    \toprule
    &&&&&  \multicolumn{4}{c}{Synthetic Test Set} &  \multicolumn{4}{c}{Real Test Set MOS}\\
    \cmidrule(lr){6-9} \cmidrule(lr){10-13}
    Model & Params. &
    \multicolumn{1}{l}{$\alpha$} &  
    \multicolumn{1}{l}{$\beta$} 
        &  {Data Augmentation} 
        &  {SI-SDR} &  {SDR} &  {STOI} &  {PESQ} 
        &  {NE} &  {FE} &  {DT/E} &  {DT/O}\\
    \midrule
    NSNet \cite{Braun2021InterspeechChallenge}
    & 3.5M
        &  1 &  0 &  -- 
                        &  9.95 &  10.46 &  .84 &  1.45
                        &  3.77	&  3.73	 &  3.99	&  2.23\\
    Conv-TasNet \cite{Luo2019Conv-TasNet:Separation}
    & 5.1M
        &  1 &  0 &  -- 
                        &  7.97 &  9.38 &  .83 &  1.30
                        & 3.03	&  3.21	&  3.93	&  2.25\\
    DCCRN \cite{Hu2020DCCRN:Enhancement}
    & 3.7M
        &  1 &  0 &  -- 
                        &  5.54 &  5.65 &  .84 &  1.33
                        &  3.66	&  2.83	&  3.60	&  2.71\\
    \midrule
    cD3Net single mask
    & 352K
        &  1   & 0 & -- 
                        &  13.10&  13.60&  \b.90 &  1.77
                        &  3.94	&  4.11	&  4.16	&  2.60	\\
    \midrule
    cD3Net dual mask
    & 354K
        &  1 &  0 &  -- 
                        &  13.22 &  \b13.74 &  \b.90 &  1.80
                        &  3.94	&  4.19	&  4.12	&  3.13\\
        & &  0.75 &  0.25 &  -- 
                        &  13.00&  13.51&  \b.90 &  1.84
                        & 3.94	&  4.06	&  4.10	&  2.40\\
        & &  0.50 &  0.50 &  -- 
                        &  13.21&  13.70&  \b.91 &  1.99
                        &  3.96	&  4.28	&  4.12	&  3.29	\\
        & &  0.25 &  0.75 &  -- 
                        &  \b13.26 &  \b13.74 &  \b.91 &  \b2.05 
                        &  3.94	&  4.26	&  4.13	&  3.16	\\
        & &  0 &  1 &  -- 
                        &  9.49 &  10.74 &  \b.90 &  1.97
                        &  \b4.00&  	4.37&  	4.22&  	3.68\\
        \cmidrule{3-13}
        & &  0.25 &  0.75 
        &  farend shift 
                        &  13.23 &  \b 13.73 &  \b.91 &  2.04
                        &  \b3.99	&  4.42	&  \b4.27	&  3.89\\
        &&&&  farend scale 
                        &  13.15 &  13.67&  \b.91&  1.98
                        &  3.97	&  4.28	&  4.14	&  3.20	\\
        &&&&  farend shift and scale 
                        &  13.21 &  13.70 &  \b.91 &  \b2.06
                        &  \b3.99	&  \b4.44	&  \b4.26	&  \b4.00\\
    \bottomrule
    \end{tabularx}
    \caption{Evaluation results on the synthetic test set \cite{Peng2021ICASSPSuppression} and real test set \cite{Braun2021InterspeechChallenge}. \textit{Legends for MOS scores}: NE -- singletalk nearend MOS; FE -- singletalk farend Echo DMOS; DT/E -- doubletalk Echo DMOS; DT/O -- doubletalk Other DMOS. 
    Metric values within 0.01 of the respective top scores are considered practically equivalent.  
    }
    \label{tab:results}
\end{table*} 

\subsection{Discussion}

The results for all models evaluated are shown in \Cref{tab:results}. Note that we \textbf{do not} perform any preprocessing nor postprocessing on the test set, i.e., no delay compensation or gain adjustment. The networks are fully responsible for all processing required to perform end-to-end echo/noise suppression and speech enhancement with the test set tracks as-is.

\subsubsection{Architectures}
In both the synthetic and real test sets, cD3Net models of any configuration consistently outperformed the three baseline models across most metrics, despite only having about a tenth of the parameters of the smallest baseline model. This demonstrates the ability of cD3Net architecture to learn useful intermediate features in a very parameter-efficient manner. Interestingly, NSNet, which performed the best among the baselines on the synthetic test set for most metrics, has the simplest architecture amongst the baselines and also does not use pooling. The unexpected poor performance by DCCRN and Conv-TasNet could be due to the limited number of epochs all models are trained on, as well as inherent shortcomings in their architectures. DCCRN performs frequency downsampling in each layer, while Conv-TasNet uses a basis kernel in lieu of STFT \cite{Luo2019Conv-TasNet:Separation}. The small 16-sample time-domain kernel is likely insufficient for learning effective representations for echo cancellation due to the inherent time offsets between the nearend and loopback signals.

\subsubsection{Masking techniques}
The dual-mask variant of cD3Net shows a slight improvement on the synthetic test set results compared to the single-mask variant. Although the real test set results are comparable for nearend MOS, farend singletalk Echo DMOS, and doubletalk Echo MOS, the doubletalk Other DMOS shows a significant improvement with the dual-mask technique. This is somewhat expected by design, as the single-mask technique is unlikely to be able to cope with both echo/noise suppression and speech enhancement in a doubletalk scenario.

\subsubsection{Loss function}
The models discussed so far were all trained purely using the negative SDR loss. Using a pure PMSQE loss on the dual-mask cD3Net, we found that the SDR and SI-SDR metrics on the synthetic test set degraded considerably, despite the real test set MOS indicating otherwise. Due to the lack of ground truth data on the real test set, it is not possible to exactly analyze the cause of this discrepancy, although we suspect that PMSQE may simply teach the model to optimize for perceived speech quality without regard to the actual amount of echo or noise suppressed. 

Regardless, we found that using a weighted composite loss of negative SDR and PMSQE results in a better trade-off between the SDR-based metrics and the perceptual proxies, and even improves STOI and PESQ with the 1:3 and 1:1 SDR-to-PMSQE weightings. 

\subsubsection{Data Augmentation}

Using the 1:3 weighting, which performed best on the synthetic test set, we experimented with data augmentation in the form of time shifting and amplitude scaling on the loopback signal, as described in \Cref{ssec:daug}. It is important to note that the echo and the loopback signals in the synthetic test set are nearly exactly time-aligned. On the other hand, the time offsets between the loopback and the echo signals in the real test set can be much more significant. Since we are using end-to-end models without offset compensation, it is not surprising that adding random time shifts to the farend loopback signal considerably improved the model performance on the real test set. The slight drop in the performance on the synthetic test set is likely due to the domain shift introduced by the augmentation.

Although the introduction of random farend amplitude scaling alone worsens the performance across both test sets, when used with random shifting, amplitude scaling further improved the DMOS performance in the real test set, achieving the best performances across the MOS metrics.

\section{Conclusion}

In this work, we proposed cD3Net, an end-to-end pooling-free network for joint echo cancellation, noise suppression, and speech enhancement, using a complex-valued extension of the D3Net building block, with a very small parameter count of only 354K. The complex-valued extension eliminates the need for additional linear filtering or preprocessing modules while the multidilated building blocks allow feature extraction using large receptive fields without pooling, preserving full feature resolution throughout the network. To fully utilize the farend loopback signal information, we also introduced a time-frequency domain enhancement technique using dual masks for joint AEC, DNS, and speech enhancement. Using a mixed loss training with scale-dependent SDR and PMSQE to account for enhancement in both energy-based and perceptual senses, evaluation on both synthetic and real test sets showed promising results. The proposed model achieved upwards of \SI{13.2}{\decibel} for SI-SDR, \SI{13.7}{\decibel} for SDR, and upwards of \num{0.91} for STOI, as well as scored \num{4} or above for all deep proxies of degradation MOS.

\renewcommand{\bibsection}{\section{References}}
\bibliographystyle{IEEEbib}
\bibliography{references}

\end{document}